\documentclass[preprint,3p,compress]{elsarticle}
\journal{TRIBINT}
\usepackage{graphicx}           
\usepackage{amssymb,amsmath}
\usepackage{xspace}
\usepackage{color}
\usepackage{listings}
\usepackage{colortbl}
\usepackage{subcaption}
\usepackage{tikzinclude}
\usepackage{rotating}
\usepackage{longtable}
\usepackage[percent]{overpic}
\usepackage{enumitem}
\usepackage{mathtools}
\usepackage{pifont}
\usepackage{marvosym}

\usetikzlibrary{automata,positioning}
\usetikzlibrary{shapes.multipart}
\usetikzlibrary{shapes.geometric, arrows}
\usetikzlibrary{decorations.pathreplacing}
\newcommand{\nc}{\newcommand}
\nc{\rnc}{\renewcommand}
\nc{\bs}{\boldsymbol}
\rnc{\vector}[1]{\matrix{c}{#1}}
\nc{\mm}[1]{\boldsymbol{#1}}
\nc{\mms}[1]{\boldsymbol{#1}}
\nc{\real}[1]{\Re\lbrace #1 \rbrace}
\nc{\imag}[1]{\Im\lbrace #1 \rbrace}
\nc{\dd}{\mathrm{d}}
\nc{\ii}{\mathrm{i}}
\nc{\ee}{\mathrm{e}}
\nc{\inv}{^{-1}} 
\nc{\herm}{^{\mathrm H}}
\nc{\tra}{^{\mathrm T}}
\nc{\conj}[1]{ \overline{#1} }
\nc{\normal}{\mathrm n}
\nc{\tangential}{\mathrm t}
\nc{\kn}{{k_{\normal}}}
\nc{\kt}{{k_{\tangential}}}
\nc{\ie}{i.\,e.\xspace}
\nc{\eg}{e.\,g.\xspace}
\nc{\vs}{vs.\xspace}
\nc{\cf}{cf.\,\xspace}
\nc{\myquote}[1]{`#1'}
\nc{\HB}{HB\xspace}
\nc{\etal}{et al.\xspace}
\nc{\mrm}{\mathrm}
\nc{\mum}{$µ$\mrm{m}}
\nc{\fabstand}{\,}
\nc{\fp}{\fabstand.}
\nc{\fk}{\fabstand,}
\nc{\tab}[5][tbh]{\begin{table}[#1]\centering\caption{#4\label{tab:#5}}\begin{tabular}{#2}\hline #3 \\ \hline\end{tabular}\end{table}}
\nc{\ns}{N_\text{s}}
\nc{\fig}[4][tbh]{
\begin{figure}
    \centering
    \includegraphics[width=#4\linewidth]{#2}
    \caption{#3}
    \label{fig:#2}
\end{figure}
}
\nc{\e}[2]{\begin{equation} #1 \label {eq:#2} \end{equation}}
\nc{\est}[1]{\begin{equation*} #1 \end{equation*}}
\nc{\ea}[1]{
\begin{eqnarray}
#1 \end{eqnarray}}
\nc{\east}[1]{
\begin{eqnarray*}
#1 \end{eqnarray*}}
\nc{\fref}[1]{{Fig.~\ref{fig:#1}}}
\nc{\frefo}[1]{{\ref{fig:#1}}}
\nc{\frefs}[1]{{Figs.~\ref{fig:#1}}}
\nc{\tref}[1]{{Tab.~\ref{tab:#1}}}
\nc{\trefo}[1]{{\ref{tab:#1}}}
\nc{\trefs}[1]{{Tab.~\ref{tab:#1}}}
\nc{\eref}[1]{{Eq.~(\ref{eq:#1})}}
\nc{\erefo}[1]{(\ref{eq:#1})}
\nc{\erefs}[1]{{Eqs.~(\ref{eq:#1})}}
\nc{\sref}[1]{{Section~\ref{sec:#1}}}
\nc{\srefo}[1]{\ref{sec:#1}}
\nc{\srefs}[1]{{Sections~\ref{sec:#1}}}
\nc{\ssref}[1]{{Section~\ref{sec:#1}}}
\nc{\ssrefo}[1]{\ref{sec:#1}}
\nc{\ssrefs}[1]{{Sections~\ref{sec:#1}}}
\nc{\aref}[1]{{{\ref{asec:#1}}}}
\nc{\arefo}[1]{{\ref{asec:#1}}}
\nc{\arefs}[1]{{{Appendices~\ref{asec:#1}}}}
\nc{\inst}[1]{$^{#1}$}
\nc{\naft}{N}
\nc{\ndof}{N_{\mathrm{DOF}}}
\nc{\Neq}{N_{\mathrm{eq}}}
\nc{\nfact}{N_{\mathrm{fact}}}
\nc{\nsolpt}{N_{\mathrm{pt}}}
\nc{\nord}{P}
\nc{\nnewtavg}{\overline{N}_{\mathrm{newt}}}
\nc{\regeps}{\varepsilon_{\mathrm{reg}}}
\nc{\epstol}{\varepsilon_{\mathrm{tol}}}
\nc{\sign}{\operatorname{sgn}}
\nc{\BMT}{BMT\xspace}
\nc{\FT}{T}
\nc{\Fpiezo}{F_{x}}
\nc{\gT}{g_{\mrm{T}}}
\nc{\cxbulk}{c_{x,\mrm{sample}}}
\nc{\kxbulk}{k_{x,\mrm{sample}}}
\nc{\FN}{F_{\mrm{N}}}
\nc{\FDMS}{F_{z}}
\nc{\gN}{g_{\mrm{N}}}
\nc{\ufive}{u_{z}}
\nc{\krz}{k_{z,\mrm{lo}}}
\nc{\kazp}{k_{z,\mrm{sample+up}}}
\nc{\cazp}{c_{z,\mrm{sample+up}}}
\nc{\kazcyl}{k_{z,\mrm{cylinder+up}}}
\nc{\cazcyl}{c_{z,\mrm{cylinder+up}}}
\nc{\czcyl}{c_{z,\mrm{cylinder}}}
\nc{\czsample}{c_{z,\mrm{sample}}}
\nc{\pN}{p_{\mrm N}}
\nc{\pT}{p_{\mrm T}}
\nc{\email}{malte.krack@ila.uni-stuttgart.de}
\nc{\adresA}{University of Stuttgart, Institute of Aircraft Propulsion Systems, Pfaffenwaldring 6, 70569 Stuttgart, Germany}
\nc{\adresB}{MTU Aero Engines AG, Dachauer Strasse 665, 80995 Munich, Germany}

\makeindex

\begin{document}

\begin{frontmatter}
\title{The Black Metal Tribometer: high-resolution measurement of normal load-indentation curves and partial slip hysteresis cycles}

\author{Daniel Fochler$^a$}
\ead{daniel.fochler@ila.uni-stuttgart.de}
\author{Stefan Schwarz$^b$}
\ead{stefan.schwarz@mtu.de}
\author{Lukas Kohlmann$^b$}
\ead{lukas.kohlmann@mtu.de}
\author{Malte Krack$^a$\corref{cor1}}
\ead{malte.krack@ila.uni-stuttgart.de}
\cortext[cor1]{Corresponding author. Email: \email. Postal Address: \adresA}
\address{$^a$\adresA}
\address{$^b$\adresB}

\begin{abstract}
A new instrument has been designed for measuring frictional hysteresis cycles and normal load-indentation curves.
The primary purpose of the BMT is the validation of more predictive modeling approaches for damping in friction joints.
An important original feature of the BMT is its ability to measure both normal and tangential contact behavior without having to separate the samples.
Special attention was paid to alignment and smooth motion in order to permit testing nominally flat-on-flat contacts in the microslip regime.
Examples are shown for finely resolved, undistorted hysteresis cycles with relative tangential displacement amplitudes in the sub-micrometer range at about 50 MPa nominal normal pressure, which is maintained well over the tangential load cycle.
\end{abstract}

\begin{keyword}
dry friction; jointed structures; partial slip; test rig
\end{keyword}

\end{frontmatter}

\section{Introduction}
To engineer mechanical systems against harmful vibrations, it is crucial to quantify their damping.
The amount of damping determines, in particular, if resonances can be survived, how prone a system is to dynamic instabilities, and how quickly cyclic loads decay after a shock-type load.
Damping can generally be due to inelastic processes within the material, interaction with fields of other physical origin (\eg air flow), and dry frictional contact interactions at the interfaces.
For most lightweight structures, dissipation within the material is negligible, and (dry) friction is the main cause of structural damping \cite{Brake.2018}.
Some interfaces are designed for gross sliding and/or liftoff.
This is the case, \eg, for friction dampers, like the under-platform dampers commonly used to mitigate vibrations of turbo-machinery blades \cite{srin1983}.
Other interfaces do not have damping but load transmission, leak tightness or alignment as primary purpose.
Common examples are contact interfaces at mechanical joints tightened by bolts or rivets.
Those contacts never reach gross sliding / liftoff during operation (before failure).
On the other hand, those contacts are not fully stuck either.
To understand this, it is important to emphasize that real technical surfaces are not flat, but feature form deviations, waviness and roughness.
Consequently, the real contact area may be a small fraction of the apparent (or nominal) contact area, even at high squeezing forces, and the contact pressure distribution is highly non-uniform, see \eg \cite{Campana.2011}.
Under vibratory loading, some parts of the contact area undergo microscopic relative motion, while the contact is still in a macroscopic stick phase.
This regime is called \emph{partial slip}, pre-sliding or microslip.
It is useful to note that while the local relative displacements within a frictional clamping/joint may be $<1~\mum$, the maximum (absolute) vibration level of the jointed structure may be much larger (away from the clamping/joint).
The effective damping ratio may reach some tenths of a percent for the critical low-frequency modes, and microslip is usually the main cause of dissipation if there are no dedicated means of vibration mitigation/dampers \cite{Brake.2018}.
\\
When the contacts undergo gross slip / complete liftoff, reasonable predictions of the nonlinear vibration behavior can be obtained, see \eg \cite{vandeVorst.1996,vandeVorst.1998,Wagg.2002,Krishna.2012,firr2008,Claeys.2016b,Pesaresi.2017,MonjarazTec.2022,Schwarz.2023}.
In that regime, the surface irregularities (in particular waviness and roughness) play only a minor role, because their length scale is small compared to the relative contact displacements, and/or because they are worn away rapidly.
Consequently, the associated finite compliance (contact stiffness) can often be neglected, and the contact model can be idealized to rigid unilateral behavior (Signorini conditions) in the normal contact direction and rigid Coulomb friction in the tangential contact plane, imposed between the nominal contact interfaces.
The key parameter of such a contact model is the friction coefficient, which can be selected from experience or identified using relatively simple tribometers.
In contrast, if the contacts undergo partial slip / partial liftoff, it is the present state knowledge that the vibration behavior is unpredictable \cite{Brake.2018}.
Here, it is common practice to calibrate phenomenological models, such as the Dahl or the Bouc-Wen or the Iwan model, to vibration results of existing system dynamics test rigs \cite{Worden.2007}.
By system dynamics test rig, we mean that a vibrating structure with friction joints is tested.
The transfer to other / not-yet-fabricated systems is quite limited.
Recent research efforts suggest that a sufficiently accurate description of the form \cite{ZareEstakhraji.2023} and/or the roughness \cite{Willner.2008b} is crucial to develop more predictive modeling approaches for partial slip / partial liftoff \cite{Brake.2021}.
An important prerequisite for developing more predictive partial slip models is a means to validate them.
The aim to characterize nominally flat-on-flat contacts in the partial slip regime was an important motivation for developing the Black Metal Tribometer (\BMT).
\\
Some test rigs involve application-oriented forms of joints and/or are to be categorized as system dynamics test rig, in the sense defined above, yet they are sufficiently instrumented to measure frictional force-displacement relations.
Examples are the lap joint resonator in \cite{Gaul1997,Eriten.2011,Eriten.2012}, 
the under-platform damper test rigs in \cite{Koh.2005,Botto.2018,Umer.2019,Gastaldi.2020}, 
and the resonating beam subjected to friction and wear in \cite{Tamatam.2021}. 
The typical aim of such test rigs is the validation or identification of a whole-joint or system model.
Other test rigs are better qualified to validate or identify local frictional contact models, where the intention is that the local model can be transferred to a wide range of applications.
These test rigs are referred to as \emph{tribometers} throughout this article.
\\
An important distinguishing feature of tribometers is the form of the apparent contact.
In some cases, the apparent contact area is zero, as in the common pin-on-disk tribometer (point contact), see \eg \cite{Cabboi.2016}.
Here, the contact area becomes non-zero only under (in)elastic deformation and wear.
A benefit of nominal point contacts is that no precise alignment is needed.
The technical motivation of the present work is partial slip of conformal interfaces, \ie, contacts with non-zero nominal area.
The transfer of models from point contacts to area contacts is difficult, if not impossible.
Important reasons for this are the strong non-uniformity of the pressure distribution of nominal point contacts \cite{Sidebottom.2015}, and their higher sensitivity to individual asperities \cite{Lampaert.2004}.
For nominal point contacts \cite{Hagman.1998,Hagman.1998b} and line contacts \cite{Lampaert.2004}, tribometers have been built that are able to characterize frictional force-displacement relations in the partial slip regime.
Others were developed with a focus on gross slip, and feature a rather poor resolution in the sub-micrometer range \cite{Filippi.2004,Hoffmann.2019}.
\\
A thorough literature search was done to assess the present state of tribometry for nominally flat-on-flat contacts.
This revealed that most tribometers are designed for gross or even reciprocating sliding.
Their primary purpose is to characterize the relation between friction force and sliding distance and/or fretting damage in the form of wear or fatigue.
For instance, the tribometer at the Imperial College \cite{stan2001,Schwingshackl.2012,Fantetti.2019} is commonly used to determine stick stiffness and friction coefficient from gross slip hysteresis tests, and the evolution of those parameters with wear.
Typical sliding distances are in the range of $14-50\mum$, which is useful for modeling contacts designed for gross sliding, like those at friction dampers.
The second-generation tribometer at the Politecnico di Torino \cite{Botto.2012,Lavella.2013} has a very similar purpose, and results acquired from both tribometers have been compared comprehensively in \cite{Fantetti.2024}.
The tribometer at the University of Oxford \cite{Kartal.2011b} has been used to analyze fretting wear and/or fatigue.
The tribometers at SUPMECA Paris \cite{Dion.2013,FortesDaCruz.2015} have been used for both fretting and hysteresis identification.
Commercially available fretting test rigs do not seem to permit hysteresis measurements with sub-micrometer amplitude; the product description of the 2024 Rtec Fretting Tester, for instance, specifies an actuation range starting from $5~\mum$ \cite{Rtecinstruments.2024}.
In spite of the thorough literature search, we were unable to find evidence that any existing tribometer is able to acquire undistorted hysteresis cycles of nominally flat-on-flat contacts in the partial sliding regime (sub-micrometer range).
It should be stressed that the focus of most studies is placed on the gross sliding rather than the partial sliding regime.
Many examples of partial slip hysteresis cycles can be found which appear heavily distorted.
We mention here just four examples from different research groups: Figure 3 in \cite{Fouvry.2004}, Figure 7 bottom in \cite{Dion.2013}, Figure 12 top in \cite{Lavella.2013}, Figure 16a in \cite{Fantetti.2024}.
Thanks to the merits of laser-Doppler vibrometry, the fine resolution of hysteresis cycles in the sub-micrometer range should not be a matter of instrumentation.
We are convinced that the observed distortions are caused by the presence of play, backlash and/or hysteresis within the motion concept (aside from the contact of interest).
In the design of the \BMT, an emphasis is therefore placed on ensuring very smooth motion.
\\
For the characterization of the force-displacement relation in the contact normal direction, separate indentation test rigs are commonly used, see \eg \cite{Thornley.1965,Bellow.1970}, or the more recent work by Görke and Willner \cite{Goerke.2008,Goerke.2009}.
For nominally flat-on-flat contacts, those studies show an elasto-plastic behavior, where the roughness asperities are deformed in an inelastic way during initial loading, beyond which elastic shakedown occurs.
Interestingly, friction in the partial sliding regime is also determined by the behavior of the surface roughness.
Further, the frictional properties, including friction coefficient and contact stiffness, are known to depend on the normal pressure.
An important original feature of the \BMT is its ability to acquire both normal load-indentation curves and frictional hysteresis cycles, in one run, without having to separate the samples, and to obtain useful data already from the \emph{pristine} state of the surface.
This is regarded as crucial for the validation of more predictive modeling approaches, which account for the relevant features of the surface topography (including roughness), and the effect of inelastic squeezing on friction.
Some tribological tests have been reported, where the normal load varies along the tangential load cycle.
This was either due to form deviations \cite{Hintikka.2016,Zheng.2024}, or due to the natural rig dynamics \cite{AlBender.2012}.
Finally, the tribometers in \cite{vanPeteghem.2011,Fouvry.2017,Gao.2024} should be mentioned.
These permit to vary normal and tangential load independently.
Yet they have no means to determine the \emph{relative displacement in the contact normal direction}, and published results are limited to nominal point contacts.
\\
In real joints, vibratory loads and relative displacements do not occur in the purely tangential or normal contact direction.
Under bending-type vibrations, for instance, a rolling-type combination of frictional-unilateral interactions is expected.
To predict this, it is crucial to experimentally characterize not only the tangential but also the normal load-displacement relation.
This is the purpose of the \BMT presented in this work.
A focus is placed on nominally flat-on-flat contacts in the partial sliding (sub-micrometer) regime.
In \sref{design}, the design of the \BMT is presented.
In \sref{ID}, the identification of the contact forces and displacements, both in the tangential and in the normal direction, is described, which includes the consideration of the bulk deformation.
Exemplary measurement results are shown in \sref{results}.
This article ends with conclusions and an outlook in \sref{conclusions}.

\section{Design of the tribometer\label{sec:design}}
\begin{figure}[h!]%
  \centering
  \subfloat[][]{\includegraphics[width=0.45\textwidth]{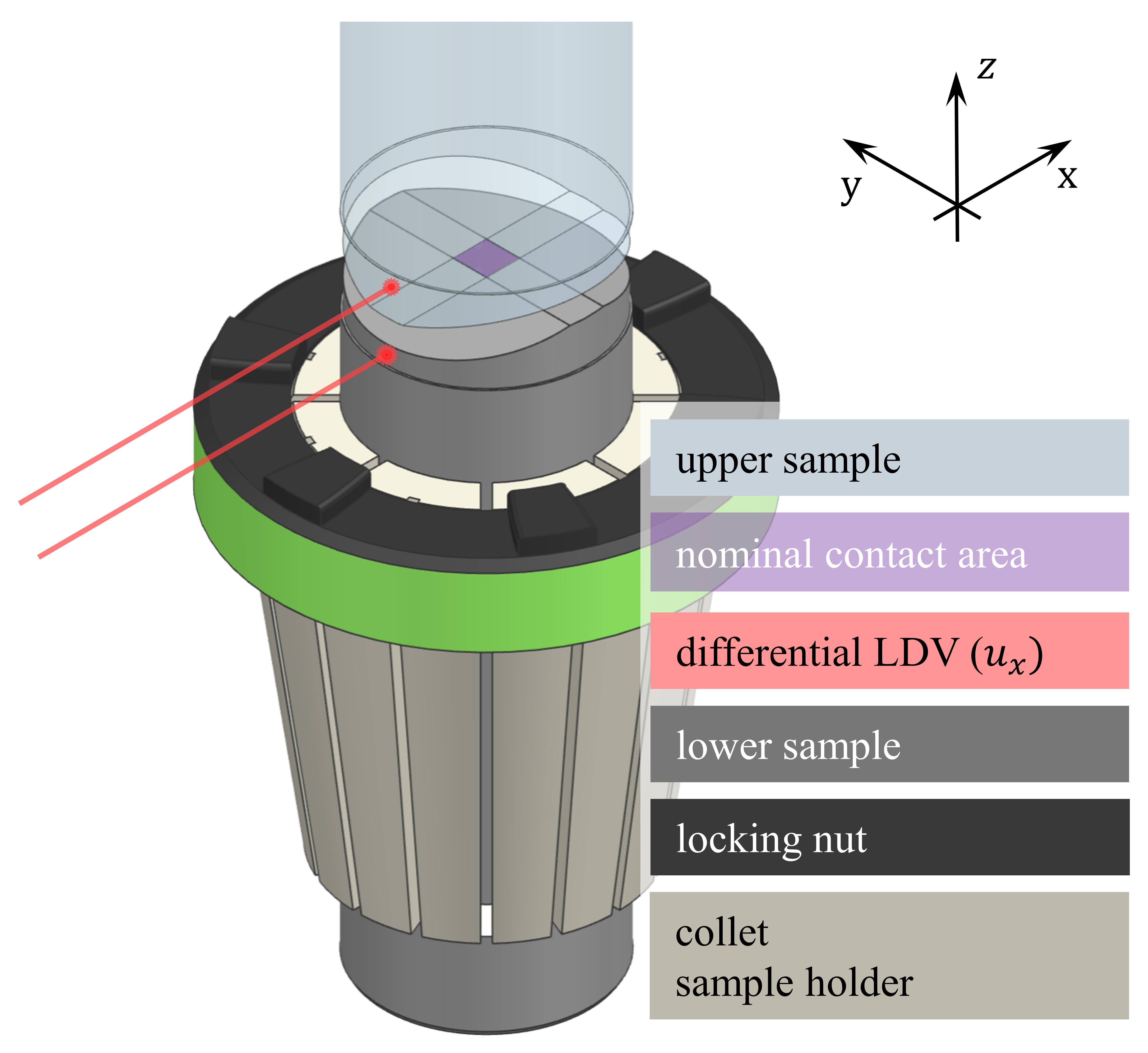}}%
  \subfloat[][]{\includegraphics[width=0.55\textwidth]{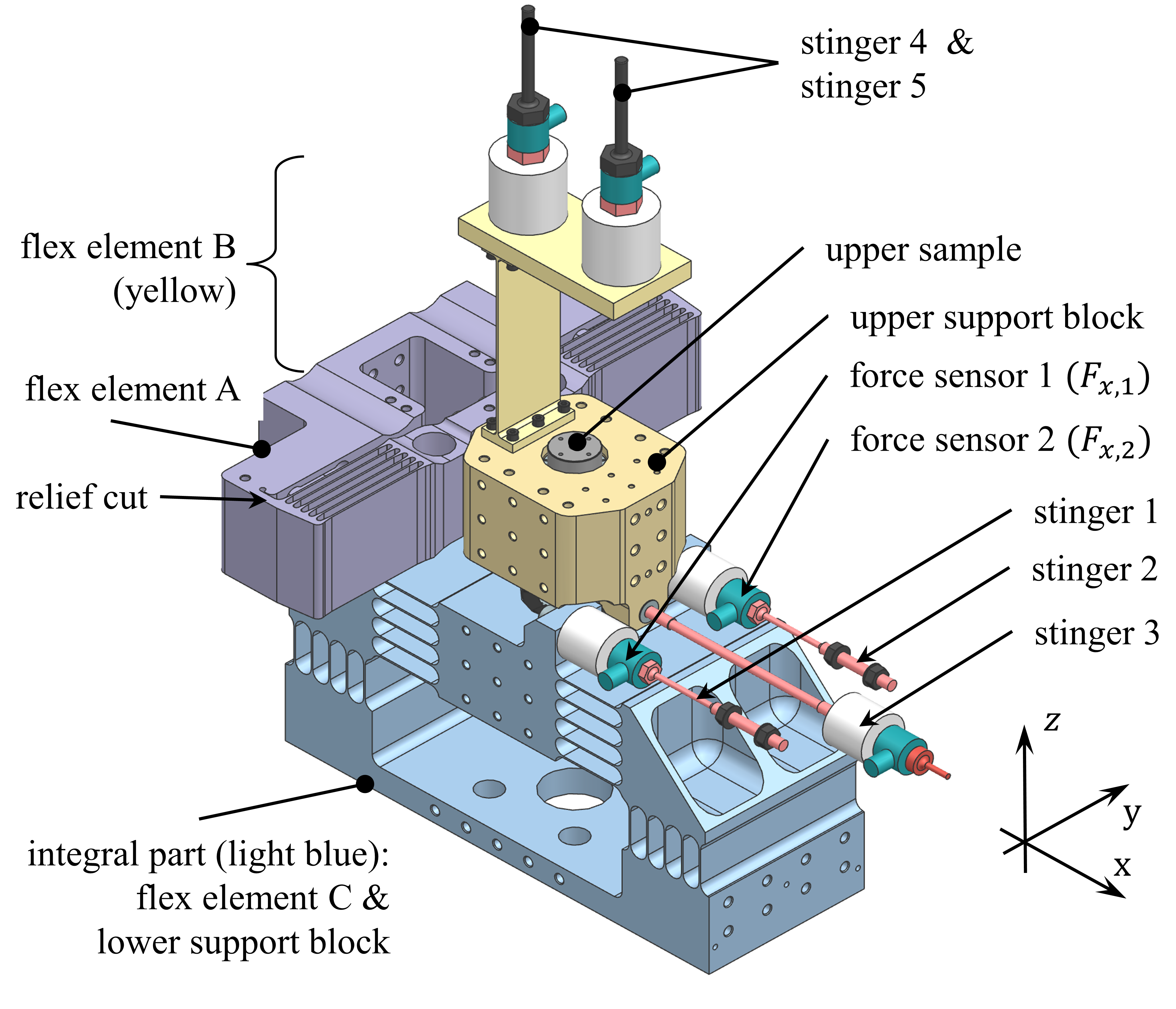}}%
  \caption{Schematics of the \BMT: (a) specimen in holder; (b) core assembly. The lower sample is only visible in (a), as it is shadowed by the upper support block in the view (b).
  } %
  \label{fig:Schematics_of_BMT}
\end{figure}
The requirement to characterize nominally flat-on-flat contacts in the partial slip regime is associated with a number of engineering challenges.
In particular, very smooth motion must be ensured, and distortions of any kind (\eg, natural dynamics, parasitic nonlinear behavior) must be avoided.
Further, a well-defined individual actuation of normal and tangential motion is required.
Finally, the alignment of a pair of nominally flat surfaces is a challenge by itself.
The presentation of the \BMT design is organized from the inside out.
First, the samples and their holders are described.
Then, the motion concept is presented, which is based on a compliant mechanism.
This is followed by the frame and aspects of isolation (vibration, thermal, electrical).
Finally, sensors and actuators are specified.
Throughout the design procedure, the principles of precision engineering were closely followed \cite{Nakazawa.1994}.

\subsection{Samples and holder\label{sec:SamplesAndHolder}}
The nominal contact area is a $1~\mrm{mm}\,\times\,1~\mrm{mm}$ square patch, obtained by arranging two beveled (angle $15^\circ$ towards tangential contact plane) samples in an orthogonal way, as illustrated in \fref{Schematics_of_BMT}a.
This is similar to the nominal contact areas tested in the Imperial College tribometer \cite{Schwingshackl.2012}.
Tight tolerances have been specified to mitigate angular misalignment, with respect to both the $x$- and the $y$-axis, of the two nominally planar surfaces.
Further, the nominal contact area is small compared to the remaining dimensions of the test rig, and thus has a relatively small rotational stiffness.
Together with the finite rotation stiffness of the compliant mechanism (described in \ssref{compliant}), this is intended to permit the compensation of the remaining angular misalignment.
\\
Apart from the region near the apparent contact area (beveled section), the samples are cylinders with circular cross section.
This way, it is relatively easy and inexpensive to produce a batch of many samples from the same original mold, in order to obtain statistically significant results.
Each of the two samples is clamped via a collet chuck (\fref{Schematics_of_BMT}a).
Those are commonly used as tool holders \eg in milling machines, and have the benefit that the samples can be easily exchanged, and fixed in a repeatable way using a torque wrench.
The outer geometry of the sample holders is conical, which ensures self-locking within the support blocks.
Relatively massive support blocks are used, to provide stiff and well-defined boundary conditions for the attached parts.
Removable guiding pins are used to achieve a repeatable angular alignment of the samples with respect to the $z$-axis.
We estimate the reassembly-induced alignment variability as $<~1^\circ$, taking into account the small play due to the guiding pins.

\subsection{Compliant mechanism\label{sec:compliant}}
To move the support blocks containing the samples, a compliant mechanism was designed.
The design concept is that the dynamic tangential load is applied to the upper support block, which thus moves along the $x$-axis (aligned with contact tangential direction),
whereas the quasi-static normal load is applied to the lower support block, which thus moves in the $z$-direction.
The compliant mechanism is implemented using flex elements (or flexures).
The \emph{upper spring} is formed by flex elements A (violet) and B (yellow), which are connected via the upper support block (also yellow) (\fref{Schematics_of_BMT}b).
It permits translation in $x$-direction (contact tangential), and constrains the remaining five rigid-body degrees of freedom.
The \emph{lower spring} is formed by flex element C (light blue) and the stingers 1 and 3, which are connected to the lower support block (also light blue; integral part with flex element C).
It permits translation in $z$-direction (contact normal), and constrains the remaining five rigid-body degrees of freedom.
\\
The stingers can be regarded as pin-type flex elements, while the remaining flex elements are of plate-type with smoothed corners \cite{Lobontiu.2002}.
Compliant mechanisms based on flex elements are popular in precision engineering, because they are practically free of play, backlash and hysteresis.
This is particularly crucial for smooth motion in the sub-micrometer range.
The plate-type flex elements have the advantage of relatively low stress.
For a single plate-type flex element, in good approximation, stiffness is proportional to the thickness, and the maximum stress is reciprocal to the square of the thickness \cite{Lobontiu.2002}.
A parallel arrangement of multiple plate-type flex elements, as used at multiple locations within the \BMT design, provides the required stiffness with a favorable stress distribution.
Additionally, such a parallel arrangement increases substantially the torsion stiffness compared to using only a single flex element.
The \BMT has a relatively high stiffness with respect to all rigid-body rotations of the samples.
On the one hand, this is regarded as crucial for achieving a well-defined motion, and, in particular, to maintain a constant normal load along the tangential load cycle.
This is also important to avoid unwanted natural dynamics of the test rig.
Yet, some remaining compliance is intended for compensating angular misalignment of the nominally planar contact areas (\cf \ssref{SamplesAndHolder}).
\\
One drawback of compliant mechanisms is limited movement.
It was ensured that the upper spring can endure $250~\mum$ tangential motion amplitude at the contact without high cycle fatigue.
This is not needed for partial sliding, but is intended to facilitate gross slip characterization in future work.
\\
Compliant mechanisms are prone to geometric nonlinearity.
In particular, clamped ends of plate-type flex elements may cause severe bending-stretching coupling.
This was found relevant in a preliminary design stage of flex element A in \fref{Schematics_of_BMT}b.
To avoid this, the indicated relief cuts were introduced, and negligible nonlinearity was verified using high-fidelity finite element analysis, up to the above specified fatigue limit.
\\
The design of the ($x$-direction) stiffness of the upper spring is crucial.
It must not be too high, in order to ensure that the applied force goes mainly into the contact rather than the elastic stressing of the upper spring.
It should also be sufficiently low to facilitate the design of a good controller.
The symmetries of the individual parts and their arrangement are regarded as important for the intended well-defined motion.
It was also ensured that the contact and the three stingers are located in one plane (\fref{Schematics_of_BMT}b).
\\
It is useful to note that flex element C (light blue) in \fref{Schematics_of_BMT}b is compliant also in the $x$-direction, thanks to the vertically oriented plate-type elements.
This is to ensure that most force goes directly into the transducers supported by stingers 1 and 2.
\\
As stated in the introduction, the natural dynamics of the rig can, in principle, distort the hysteresis measurements, and lead to a considerable variation of the normal load along the cycle.
This is particularly important in the gross slip regime, as the macroscopic stick-slip transitions have the potential to activate a broad band of modal frequencies.
The present work is limited to the partial slip regime, and it will be shown in \sref{results} that the measurements are free from distortions that could have been caused by natural rig dynamics.
Consequently, a detailed presentation of modal analysis results is regarded as beyond the scope of this paper.
Yet, it should be remarked that it was verified by finite element analysis that undesired modes have frequencies at least 5 times higher than the operating frequency, both under tied and free siding contact conditions.
\subsection{Frame and isolation\label{sec:FrameAndIsolation}}
\begin{figure}
    \centering
    \includegraphics[width=0.90\linewidth]{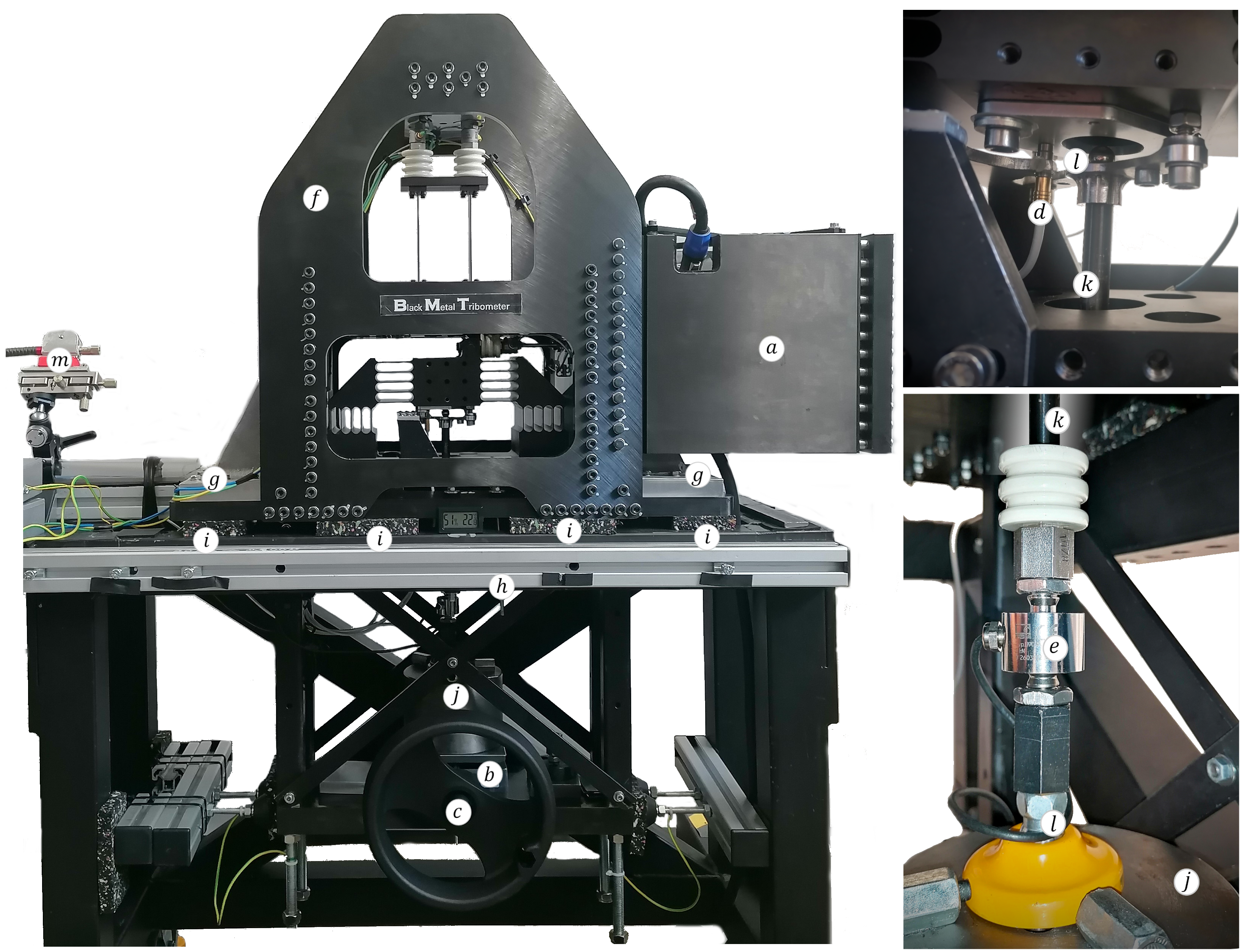}
    \caption{Photo of \BMT: (left) overview; (right) details showing sensors in normal load path.
    \textit{a}) shaker backpack; \textit{b}) machine mount; \textit{c}) hand-wheel; \textit{d}) capacitive displacement sensor ($\ufive$); \textit{e}) strain-gauge-based force sensor ($\FDMS$) \textit{f}) front plate \textit{g}) (left/right) aluminum angle plate; \textit{h}) table; \textit{i}) rubber pads; \textit{j}) initial weight resting on machine mount; \textit{k}) push rod (in normal load path); \textit{l}) spherical hinges; \textit{m}) laser heads.
    }
    \label{fig:Photos_of_BMT}
\end{figure}
%
A photo of the \BMT is shown in \fref{Photos_of_BMT}.
The compliant mechanism is stiffly supported by a thick base plate, a symmetric arrangement of two large aluminum angle plates from the left and the right, thick A-shaped front and rear plates, and a stiff top bar.
With the exception of the aluminum angle plates, all load-bearing parts are made from various high-quality strength (\fref{Photos_of_BMT} black) or stainless steels (\fref{Photos_of_BMT} shiny).
Some bolted connections are inevitable to ensure manufacturability, exchangeability  of components and to incorporate force transducers into the load path.
Those connections should have negligible effect on the system's nonlinearity, \eg in the form of play, backlash or hysteresis, and non-repeatability, \eg due to wear.
To a large extent, this is achieved by the compliant mechanism, which ensures that the strain is concentrated in the flex elements, instead of in the contact regions of the bolted connections.
In addition, high, uniformly distributed contact pressures in the bolted connections were ensured by design.
\\
The vibration isolation is achieved via soft rubber granulate pads mounted between base plate and table, and between table and floor.
Electrical and heat insulation was implemented, using, among others, the ceramic parts attached near the force transducers (\fref{Schematics_of_BMT}b, \fref{Photos_of_BMT}-right-bottom).
That insulation is not relevant in the present work, but could be useful for electrical contact resistance tests and/or the implementation of a heating module in the future.
\\
The reason for the name \emph{Black Metal Tribometer} becomes apparent from the photo in \fref{Photos_of_BMT}.
The black color originates from the chosen anti-corrosive surface finish.
A galvanic burnishing was used, which creates a very thin black coat of iron oxides.
The benefits of galvanic burnishing are the relatively low process temperatures and the low thickness of the coating.
Both aspects ensure negligible effects on mechanical properties and geometric tolerances (unlike the commonly uses lacquer).

\subsection{Instrumentation and signal processing \label{sec:instrumentation}}
For the tangential actuation, an electro-dynamic exciter (shaker) was applied via a stinger to the upper spring.
The shaker is stiffly supported by the backpack-type construction indicated in \fref{Photos_of_BMT}.
The stiff support is needed for quasi-static (non-resonant) operation, since otherwise the shaker would just push itself back instead of deforming the spring.
A sinusoidal voltage is fed to the shaker amplifier with a fixed nominal operating frequency of $40~\mrm{Hz}$.
The operating frequency should be sufficiently low to stay far away from any undesired natural frequency.
On the other hand, it should be sufficiently high to observe a significant number of cycles in a reasonable time.
The coincidence with the German power grid frequency of $50~\mrm{Hz}$ should of course be avoided.
As shown in \sref{results}, the selected operating frequency led to smooth motion, free from potential distortions by undesired natural rig dynamics.
Although it is well-known that dry friction is in very good approximation rate-independent, see \eg \cite{Lampaert.2004}, a reasonably high operating frequency is convenient to analyze wear in future work.
Feedback control is used to adjust the voltage amplitude until a desired tangential displacement amplitude is reached.
\\
For the normal actuation, a wedge-type precision machine mount is used (\fref{Photos_of_BMT}-\emph{b}), which applies a force to the lower sample via a push rod (\fref{Photos_of_BMT}-\emph{k}) and two spherical hinges (\fref{Photos_of_BMT}-\textit{l}).
The upper spherical hinge is simply implemented using a ball from a ball bearing.
A weight (\fref{Photos_of_BMT}-\emph{j}) rests on the machine mount to ensure compressive loading and to prevent play.
The machine mount is operated manually via a hand wheel (\fref{Photos_of_BMT}-\emph{c}).
Consequently, wear and inelastic processes within the load path will reduce the contact normal load.
An upgrade to a feedback-controlled operation, which maintains a constant contact normal load, is planned.
\\
Piezo-electric force transducers are attached to the lower support block (\fref{Schematics_of_BMT}).
They are mounted via stingers to ensure that only the dynamic force in the $x$-direction is acquired, and negligible bending moments are transmitted. 
A differential laser-Doppler vibrometer (LDV) with a displacement decoder is used to acquire the relative motion in the $x$-direction between the points indicated in \fref{Schematics_of_BMT}a.
Reflective tape is attached to the samples at the respective locations to improve the signal-to-noise ratio and, in particular, to prevent signal dropouts.
A strain-gauge-based force sensor, which captures both static and dynamic signal content, is placed between spherical hinges in the normal load path below the lower spring (\fref{Photos_of_BMT}-\emph{e}).
A capacitive displacement sensor is used to measure the motion between a sensor holder attached to the thick base plate (kinematic reference) and the lower support block (\fref{Photos_of_BMT}-\emph{d}).
The force transducers at flex element B are used for contact detection.
\\
A dSPACE MicroLabBox is used for data acquisition and feedback control with a sampling frequency of $10~\mrm{kHz}$.
When normal load-indentation curves are determined, a low pass filter with a cutoff frequency of $400~\mrm{Hz}$ is applied to the data acquired with the capacitive displacement sensor and the strain-gauge-based force sensor.
This seems justified given that the normal actuation is quasi-static and, hence, not expected to trigger natural rig dynamics.
The described low pass filter is needed to obtain smooth normal load-indentation curves.
The tangential actuation, on the other hand, occurs at the above specified operating frequency of $40~\mrm{Hz}$ and has the potential to trigger natural rig dynamics.
To verify smoothness of the tangential motion, thus, no low pass filter should be applied.
However, both the piezo-electric force transducers, and the laser-Doppler vibrometer with displacement decoder, which acquire tangential force and displacement, respectively, are prone to drift.
To mitigate this, the mean value is determined using a relatively slow low pass filter, and this is subtracted from the data acquired with those sensors.

\section{Identification of contact forces and displacements\label{sec:ID}}
Even if the design intent was to measure the contact forces and displacements as directly as possible, it is imperative to critically address secondary load paths, as well as the contribution of inertia loads and bulk deformation.

\subsection{Tangential contact\label{sec:tangentialID}}
The tangential (friction) force $\FT$ is simply obtained as
\ea{
\FT = \Fpiezo\fk \label{eq:T}
}
where $\Fpiezo = F_{x,1}+F_{x,2}$ is the sum of the forces $F_{x,1}$ and $F_{x,2}$, acquired from force sensor 1 and 2 (\fref{Schematics_of_BMT}b).
The elastic and inertia forces generated by flex element C (\fref{Schematics_of_BMT} light blue) have been estimated to be $<1\%$, as explained in the following, and deemed negligible.
\\
The relatively stiff stingers constrain the motion of the lower spring in the $x$-direction, which mitigates both elastic and inertia forces.
In addition, the elastic forces are small because of the relatively high compliance of flex element C compared to the stiffness of the stingers, as mentioned already in \ssref{compliant}.
In fact, this high compliance of flex element C is important to achieve a favorable splitting of the tangential force flow; if flex element C was too stiff in the $x$-direction, only a small fraction of the tangential force would be seen by the force transducers ($F_{x,1}$, $F_{x,2}$).
To estimate the residual elastic and inertia forces generated by flex element C, a highly-sensitive acceleration sensor was attached to the lower support block.
Using a finite element model of the lower spring, the inertia and elastic forces were estimated.
With this, a compensation of the actual inertia and elastic forces would in principle be possible.
However, this was found to increase the noise level considerably.
Further, the sum of inertia and elastic forces was estimated to not exceed $1\%$ of the tangential contact force amplitude in the intended range of operation conditions.
Hence this compensation was not further considered.
\\
The relative displacement $\gT$ between the contact interfaces in the tangential direction is determined as
\ea{
\gT = u_x - \cxbulk \FT\fp \label{eq:gT}
}
Herein, $u_x$ is the relative displacement between the two points indicated in \fref{Schematics_of_BMT}a, where the differential LDV is applied,
and $\cxbulk$ is the $x$-direction bulk compliance of the samples between those points.
$u_x$ counts from zero when no tangential load is applied, \ie, from the static equilibrium configuration.
The bulk displacement contributed to about one quarter to $u_x$ for the frictional hysteresis loops illustrated in \sref{results}, and hence this compensation is deemed relevant.
\\
The bulk stiffness $\kxbulk = \cxbulk^{-1}$ was determined with a finite element analysis, where the apparent contact area was tied.
The Young's modulus was experimentally identified from impact hammer modal testing of a full cylinder of batch-identical steel.
Different boundary conditions were tested where rotation of the samples was either constrained or determined as a result of the finite rotational compliance provided by the upper and lower springs.
It was found that the rotation of the samples has negligible influence on $\cxbulk$.
The highest strains are located in the beveled section of the samples.
Thus, $\cxbulk$ is sensitive to the as-manufactured web thickness of the samples; it is in good approximation reciprocal to the apparent contact area.
Therefore, the surface of each sample was scanned with a white light interferometer (Zygo\texttrademark, Nexview\texttrademark NX2 3D Optical Profiler 
) to determine this parameter.
The surface scans are illustrated in \fref{surfaceScan} and discussed later.
This revealed that the average web thickness is only $0.93~\mrm{mm}$ instead of the $1~\mrm{mm}$ specified in the technical drawing. 

\subsection{Normal contact\label{sec:normalID}}
By actuating the machine mount, the initial gap between the pair of samples is closed and the normal load is applied.
The actuation of the machine mount stresses the lower spring.
The quasi-static force flow in the $z$-direction is illustrated in \fref{model}.
The force applied with the machine mount  the lower spring, as illustrated in \fref{model}.
Hence, the normal contact force $\FN$ is determined as
\ea{
\FN = \FDMS - \krz \ufive \fk \label{eq:N}
}
where $\FDMS$ is the force acquired with the strain-gauge-based sensor (\fref{Photos_of_BMT}-\emph{e}), and $\krz$ and $\ufive$ are the stiffness and the elastic displacement of the lower spring in the $z$-direction.
$\ufive$ is obtained with the capacitive sensor (\fref{Photos_of_BMT}-\emph{e}), and counts from zero when $\FDMS=0$, in order to measure only the elastic displacement, relative to the static equilibrium configuration.
$\krz$ is determined as the slope of the load-displacement relation in the range where the contact is still open, $\FN=0$, so that $\FDMS = \krz\ufive$.
Exemplary results are shown in \ssref{NormalStiffnessID}.
Note that the normal contact force identification does not rely on the force transducers at flex element B (\fref{Schematics_of_BMT}), but only on the strain-gauge-based sensor ($\FDMS$), and the capacitive displacement sensor ($\ufive$).
\fig[htb]{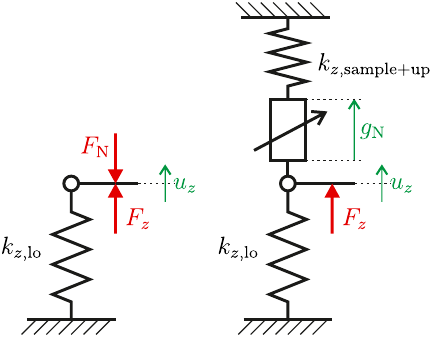}{Lumped-parameter model of compliant mechanism for the movement in $z$-direction.}{0.4} 
%
\\
The relative displacement $\gN$ between the contact interfaces in the normal direction is determined as (\fref{model})
\ea{
\gN = \ufive - \cazp\FN \fk \label{eq:gN} 
}
where $\cazp$ is the bulk compliance of the samples and the upper spring (including the sample holders).
This way, $\gN$ becomes larger when the samples approach each other.
Further, $\gN$ includes the somewhat arbitrary initial gap between the samples, \ie, $\gN$ will assume an appreciable non-zero value before the contact closes.
It is useful to note that there is no well-defined reference for setting $\gN=0$ in the case of rough surfaces (regardless of the measurement technique).
Later, a value $\Delta\gN=\gN-\gN(\FN=0)$ will be used.
\\
To determine $\cazp$, a test with a monolithic cylindrical sample (without beveled sections and without an internal contact interface) was done, which was fixed on both sides via the sample holders.
In this test, it holds that $\FDMS = (\krz + \kazcyl)\ufive$, where $\kazcyl$ is the stiffness of this monolithic cylinder, the upper spring and the sample holders.
As $\krz$ is identified as described below \eref{N}, this test yields $\kazcyl = \cazcyl^{-1}$.
The compliance $\cazp$ is then obtained as
\ea{
\cazp &=& \cazcyl + \czsample - \czcyl \fk \label{eq:cazp}
}
where $\czsample$ is the bulk compliance of the pair of samples with the beveled geometry, while $\czcyl$ is that of the corresponding section as full cylinder.
Both $\czsample$ and $\czcyl$ are obtained from a finite element analysis, analogous to the estimation of $\cxbulk$.

\subsection{Stiffness identification (without contact) and limitation of the normal displacement measurement\label{sec:NormalStiffnessID}}
\begin{figure}[h!]%
  \centering
  \subfloat[][configuration with typical sample pair before contact is closed]{\includegraphics[width=0.48\textwidth]{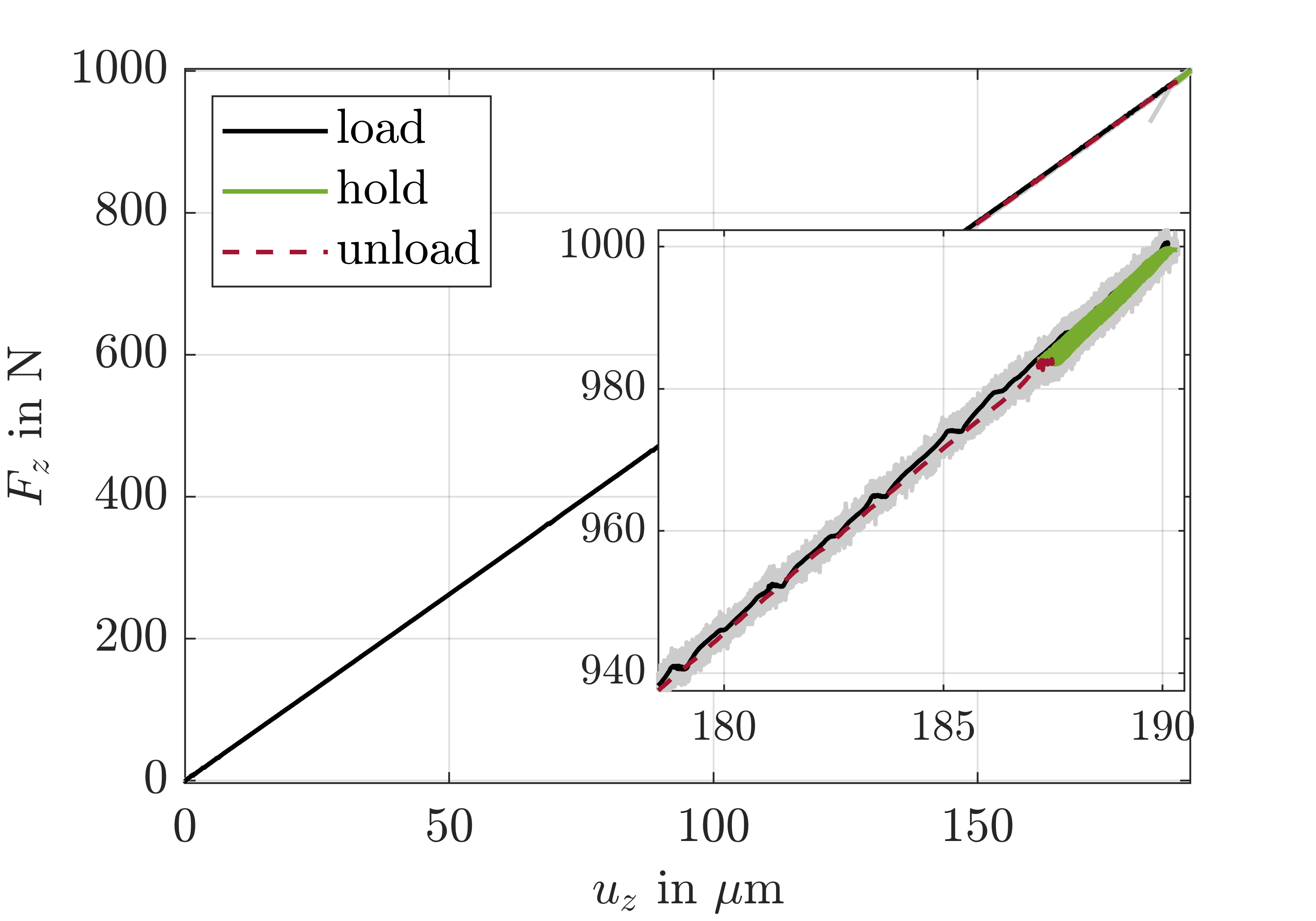}}%
  \subfloat[][configuration with with a monolithic cylinder fixed between \\ sample holders]{\includegraphics[width=0.48\textwidth]{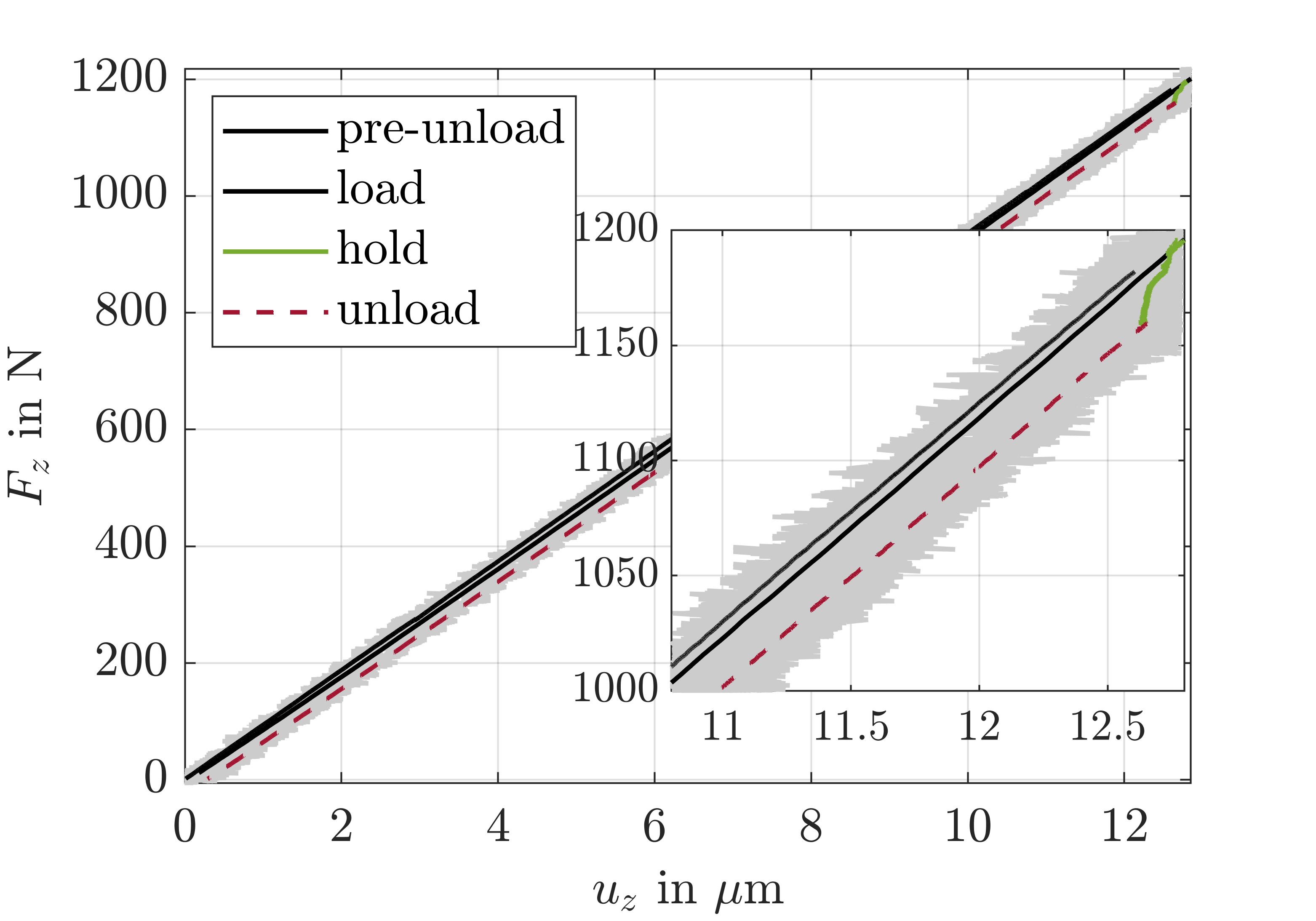}}%
  \caption{Stiffness identification of the \BMT in normal direction: $\FDMS$ vs. $\ufive$. Raw data (before application of low-pass filter specified in \ssref{instrumentation}) is shown in gray in the background.}%
  \label{fig:NormalStiffnessID}
\end{figure}
%
\fref{NormalStiffnessID}a shows the relation between $\FDMS$ and $\ufive$ in a range before the contact is closed.
Apparently, the characteristic of the lower spring is perfectly linear, and does not show any signs of hysteresis.
The extracted stiffness is $\krz = 5.3~\mrm{N}/\mum$.
The variability due to linearity is less than $\pm 1\%$, even when considering tests carried out over a measurement campaign of several weeks, and temperature differences of several Kelvin.
Thus, $\krz$ is not regarded as an important source of uncertainty.
Remarkably, however, an unloading of $15~\mrm{N}$ is observed over a hold time of $50~\mrm{min}$ ($10~\mrm{N}$ over $12~\mrm{min}$).
This appears too fast to be caused by sensor drift, especially in the initial phase of the hold time.
More importantly, the unloading takes place along the invariant characteristic of the lower spring.
The most likely cause for the observed unloading is a \emph{viscous oil extrusion process} within the lubrication of the machine mount.
As mentioned already in \ssref{instrumentation}, an upgrade of the \BMT is planned, where the normal load is maintained using feedback control.
\\
\fref{NormalStiffnessID}b shows the relation between $\FDMS$ and $\ufive$ for the configuration with a monolithic cylinder fixed between the sample holders.
In this case, the lower spring $\krz$ acts in parallel with the stiffness of the assembly comprising the upper spring, the sample holders, and the cylinder.
Consequently, the characteristic is much steeper.
The results were obtained from a test where a certain load was applied initially (pre-load, not shown for clarity of figure), followed immediate by an unloading (called \emph{pre-unload} in the legend) and repeated loading (\emph{load}).
After a holding phase (\emph{hold}), the load was removed (\emph{unload}).
Apparently, linearity and (immediate) repeatability are very good.
A stiffness $\kazcyl = 93~\mrm{N}/\mum$ has been identified, with a variability of less than $\pm 2\%$, also when considering repeated assemblies of the monolithic cylinder.
\\
As in the case where only the lower spring is loaded (\fref{NormalStiffnessID}a), we expect some unloading due to the machine mount.
As one can see in \fref{NormalStiffnessID}b, however, a more severe unloading occurs in this configuration:
It amounts to $35~\mrm{N}$ over a hold time of $50~\mrm{min}$ ($20~\mrm{N}$ over $10~\mrm{min}$).
Note that the applied load is very similar.
This suggests that the unloading is not caused by the processes within the machine mount alone.
Indeed, we attribute this to settling processes within the contacts of the sample holders.
A smooth, degressive time evolution was observed, and we refer to the resulting component of $\ufive$ as \emph{viscous displacement}.
The sub-micrometer extent of this displacement makes it hard to clarify its physical origin.
In our view, probable causes are, again, viscous oil extrusion, or creep.
In the given configuration, a viscous displacement of the upper sample holder would increase $\ufive$, whereas that of the lower sample holder would decrease $\ufive$.
In \fref{NormalStiffnessID}b, one can see a drift towards the right from the spring characteristic.
This suggests that the viscous displacement at the upper sample holder dominates in the depicted case.
\\
Note that the viscous displacement at the lower sample holder most likely occurs also in the configuration where only the lower spring is loaded (\fref{NormalStiffnessID}a).
In that configuration, however, the full load $\FDMS$ is applied to the lower spring, instead of being shared with the upper spring, and since the displacement is measured directly at the lower spring, the viscous displacement has no effect on the characteristic shown in \fref{NormalStiffnessID}a.
\\
Fortunately, the viscous displacement occurs relatively slowly, \ie, with significant changes taking place over minutes rather than seconds.
Thus, by actuating the machine mount in a reasonably quick way, say over (tens of) seconds, meaningful changes $\Delta\gN$ can be obtained using \eref{gN}.
Merely the absolute value of $\gN$ will vary in an unknown and slow way.
Recall that the absolute value of $\gN$ is meaningless anyway in lack of a suitable reference, as discussed below \eref{gN}.
An upgrade of the \BMT is planned where $\gN$ is measured in a more direct way.
It should be emphasized also that the uncertainty in $\gN$ does not affect the measured normal force $\FN$, nor has it an effect on the measured tangential force $\FT$ and displacement $\gT$.

\section{Exemplary results\label{sec:results}}
In this section, exemplary results for the normal and tangential characterization of a nominally flat-on-flat contact are shown.
A comprehensive test campaign has been carried out involving 35 sample pairs (7 configurations of different roughness and contact dimensions, 5 realizations each).
The results will be presented in a follow-up paper where the focus is placed on the validation of a contact simulation method.
To avoid redundancy, results in this article are shown only for a single sample pair.
The apparent contact area was a $1.00~\mrm{mm}~\times~1.17~\mrm{mm}$ rectangular patch.
This was determined by a surface scan as described in \ssref{tangentialID}.
The results of the surface scan are depicted in \fref{surfaceScan}.
The surface was treated using spark erosion resulting in an isotropic roughness of about $\mrm{Ra}=1.6~\mum$
With some imagination, one can infer a curvature where the nominally flat surface transitions into the beveled section (instead of the nominal kink).
Roughness appears to dominate the overall topography, and no distinct form deviation is visible.
The forces $\FN$ and $\FT$ are divided by the apparent contact area, to obtain the depicted apparent pressure $\pN$ and traction $\pT$.
\begin{figure}[h!]%
   \subfloat[][]{\includegraphics[width=0.52\textwidth]{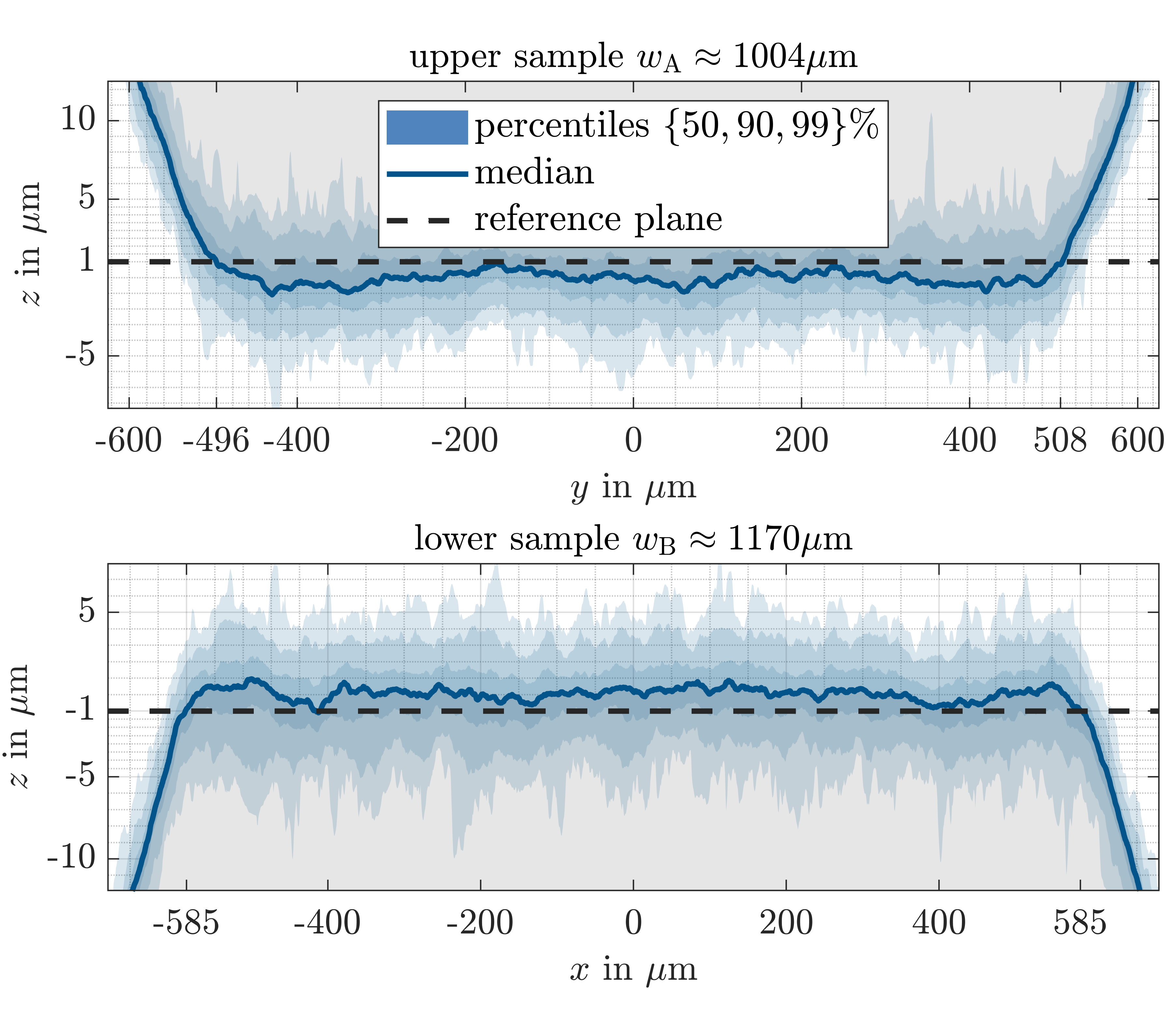}}
   \subfloat[][]{\raisebox{0.0cm}{\includegraphics[width=0.45\textwidth]{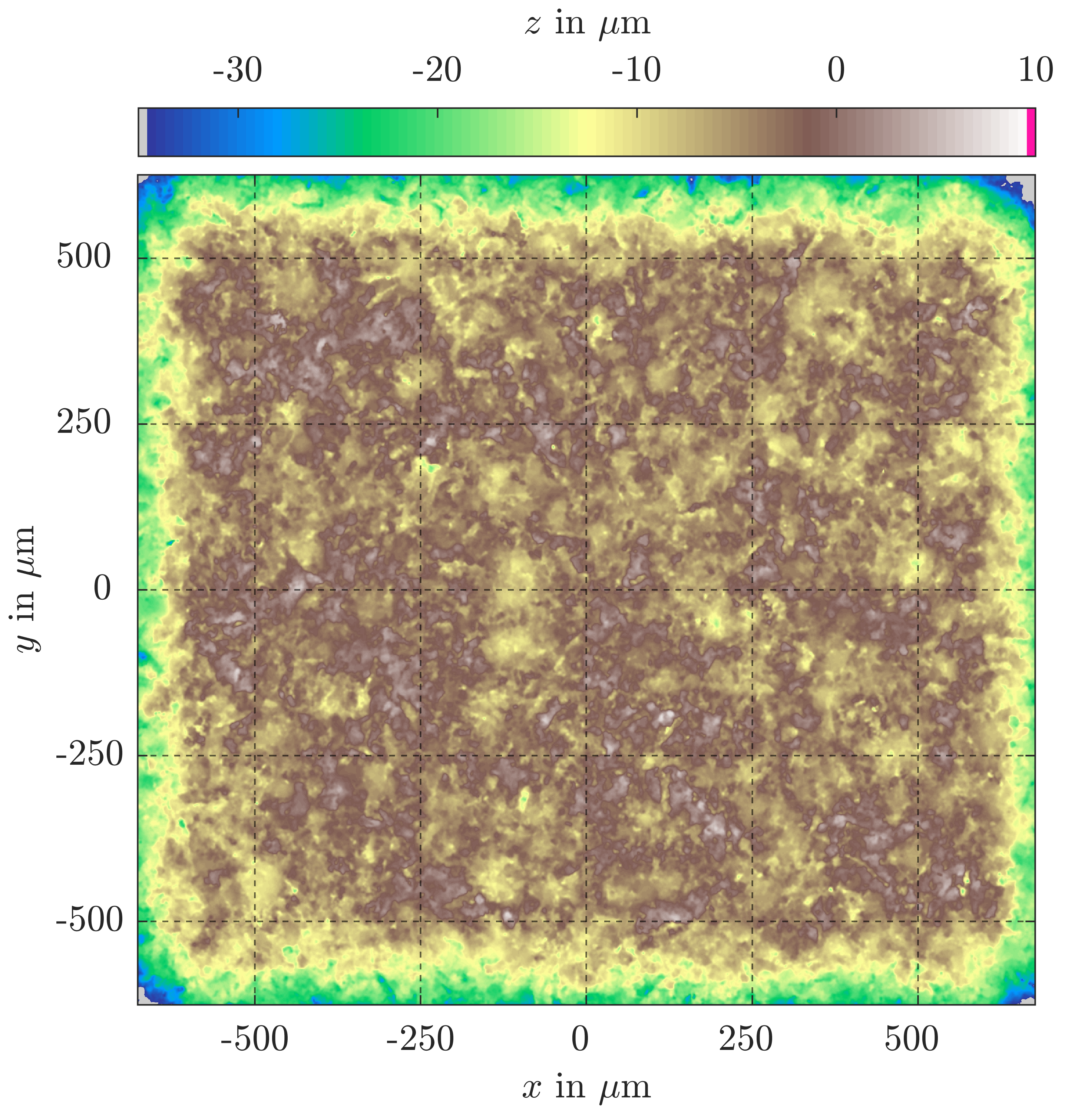}}}
  \caption{Surface scan of sample pair: (a) side view; (b) composite surface.}
\label{fig:surfaceScan}
\end{figure}
\\
The normal and tangential force-displacement relations are presented in \ssrefs{normal} and \ssrefo{tangential}, respectively.
The practically constant time evolution of the normal load over the tangential load cycle is verified in \ssref{normalLoadInvariance}.
Finally, the alignment of the nominally planar surfaces is checked in \ssref{alignment}.

\subsection{Normal force-displacement relation\label{sec:normal}}
\begin{figure}[h!]%
  \centering
  \includegraphics[width=0.5\textwidth]{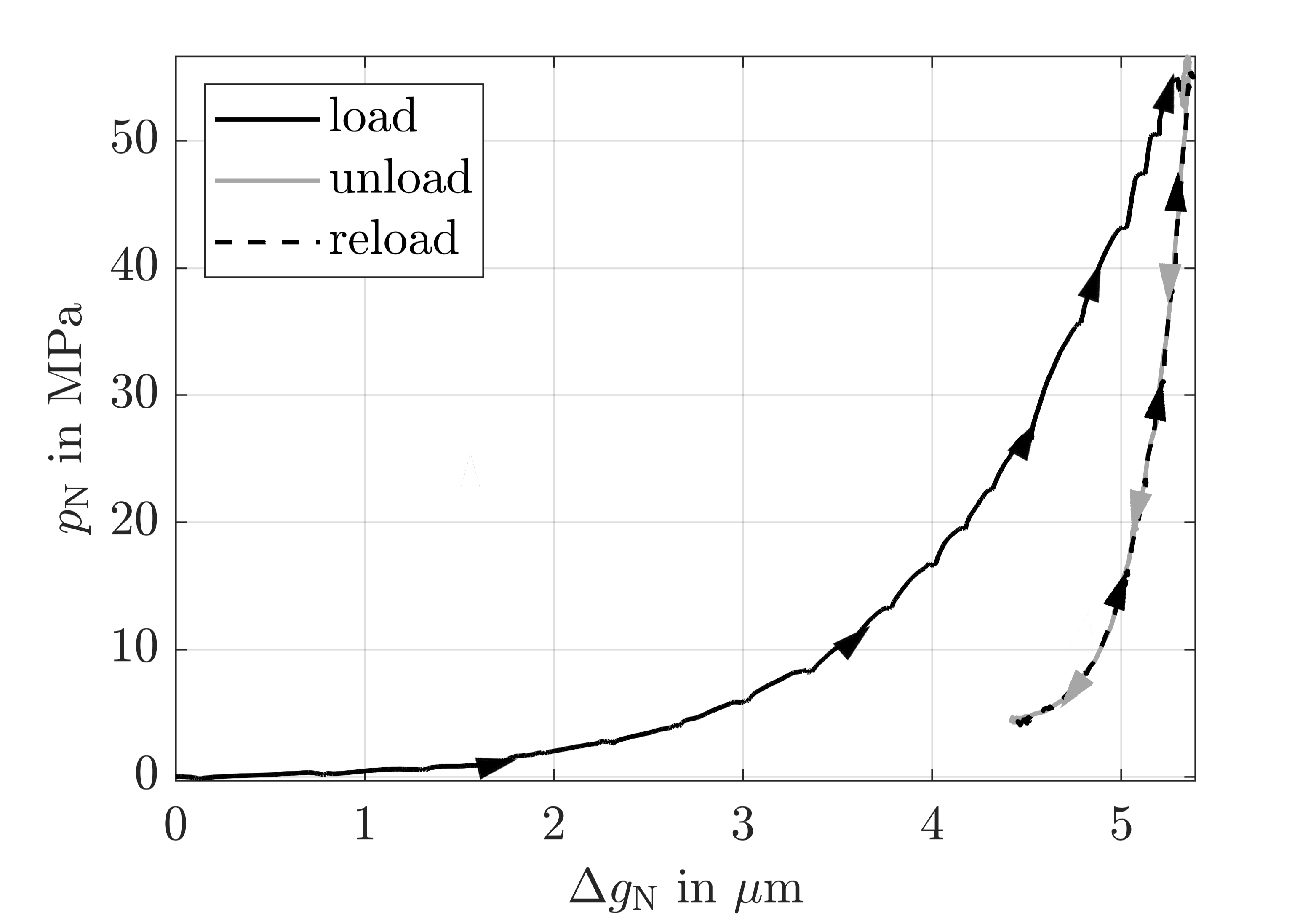}
  \caption{Normal force-displacement relation for the exemplary sample pair}
  \label{fig:NormalContact_example}
\end{figure}
\fref{NormalContact_example} shows the relation between apparent pressure $\pN$ and normal displacement.
The displacement $\Delta\gN = \gN - \gN(\pN=0)$ counts from the zero when the contact starts to close so that $\FN$ (and hence $\pN$) increases from zero.
During the first loading, some of the highest surface roughness asperities undergo plastic deformation.
Subsequently, elastic shakedown occurs as long as the load is not further increased.
Thus, unloading and reloading take place on the same curve.
After shakedown, the normal load-displacement curve resembles a power-law or exponential relation, which was verified by regression (not shown).
With regard to all those aspects, the results shown in \fref{NormalContact_example} are very similar to those obtained, for instance, by Görke and Willner \cite{Goerke.2008,Goerke.2009}.

\subsection{Tangential force-displacement relation\label{sec:tangential}}
\begin{figure}[h!]%
  \centering
  \includegraphics[width=1.0\textwidth]{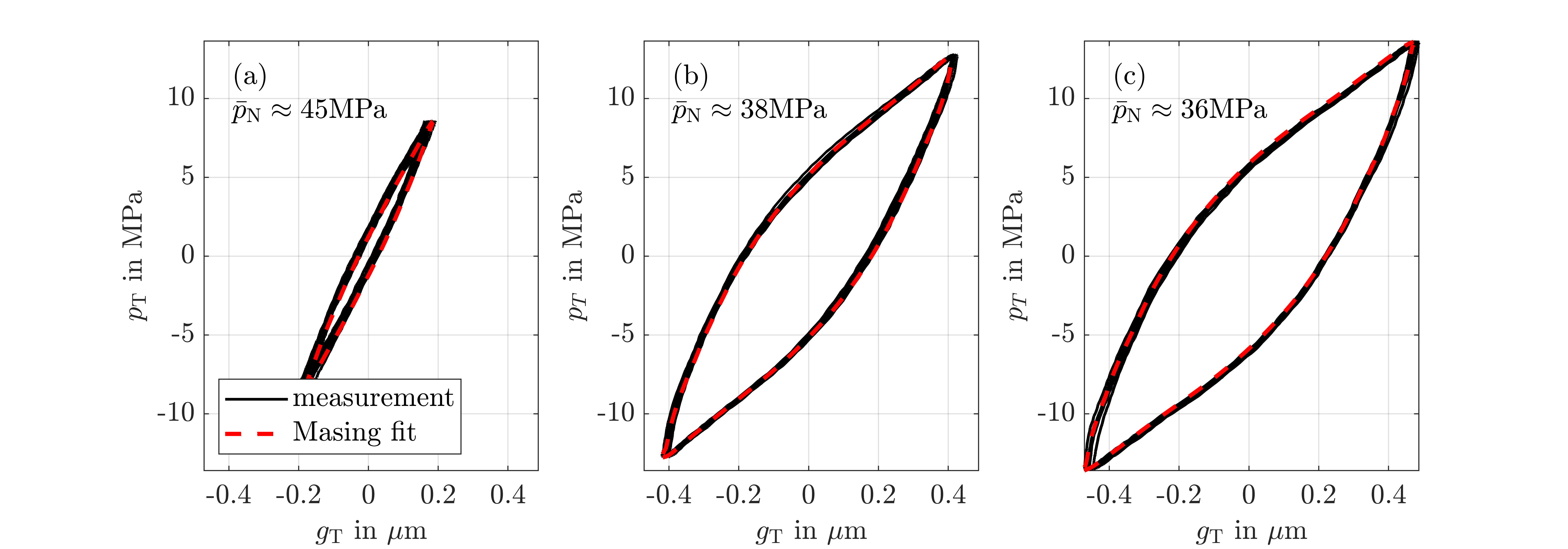}
  \caption{Tangential force-displacement relation for the exemplary sample pair: (a)-(c) show 10 tangential load cycles obtained for different tangential load levels at the specified normal pressures, acquired during the time spans highlighted in \fref{pN-time}.
  }
  \label{fig:TangentialContact_example}
\end{figure}
\fref{TangentialContact_example} shows the relation between apparent traction $\pT$ and relative displacement $\gT$ of the two contact interfaces.
As explained before, the normal load decreases slowly, due to both viscous displacements (machine mount and sample holders) and possible wear within the contact of interest.
This way, tangential load-displacement relations are obtained for different apparent normal pressures $\pN$.
At the same time, the set value for the relative tangential displacement amplitude was increased during the test.
Representative results are depicted for three different apparent normal pressures and tangential displacement amplitudes in \fref{TangentialContact_example}.
\\
When considering the depicted 10 consecutive load cycles in each sub-figure of \fref{TangentialContact_example}, one can verify that the results form very well-repeatable hysteresis cycles.
This only holds for observations over a short time window, due to the above described deliberate change of the tangential loading, and the inherent normal unloading.
From the sequence of hysteresis cycles, one can derive a backbone (also called initial loading curve) in accordance with the Masing rules, see \eg \cite{Mathis.2020}.
The hysteresis cycles reconstructed from this backbone agree well with the measured ones.
The extracted backbones themselves are not shown for clarity of the figure.
From this, one can conclude that the frictional hysteresis cycles in the partial sliding regime follow the Masing rules.
The backbone shows a softening / saturation trend, typical for dry friction.

\subsection{Verification of smooth motion and invariant normal load over tangential load cycle; \label{sec:normalLoadInvariance}}
\begin{figure}[h!]%
  \centering
  \includegraphics[width=1.0\textwidth]{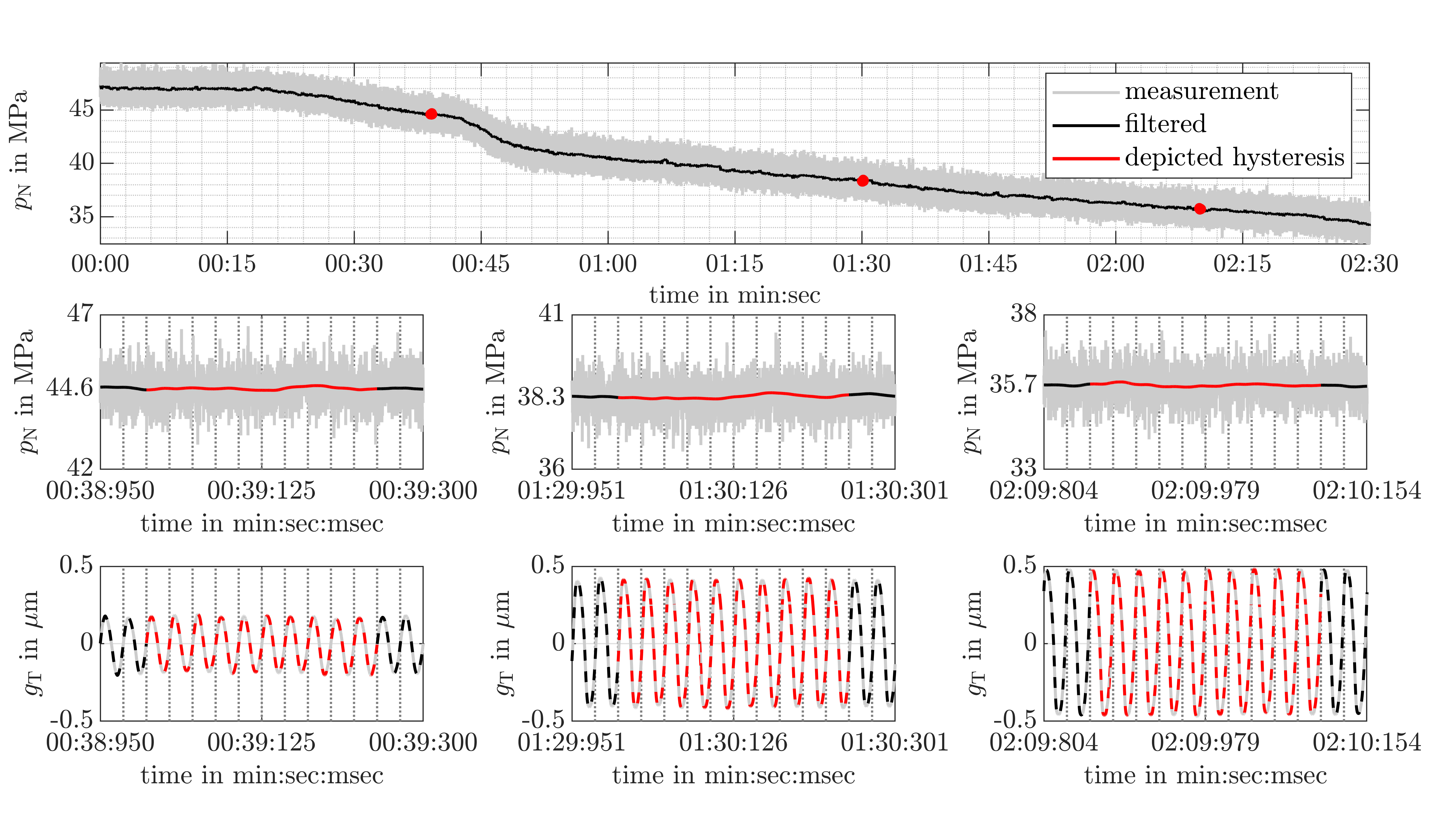}
  \caption{Evolution of normal pressure $\pN$ and relative tangential displacement $\gT$ over time: (top row) overview of $\pN$; (middle row) zooms into time spans for which the hysteresis cycles are shown in \fref{TangentialContact_example}; (bottom row) corresponding relative tangential displacement $\gT$.
  Raw data of $\pN$ (before application of low-pass filter specified in \ssref{instrumentation}) is shown in gray in the background.
  Dotted vertical lines in the zooms indicate tangential load periods.
  }
  \label{fig:pN-time}
\end{figure}
The normal unloading over time is shown in \fref{pN-time}.
When one zooms into the time frames corresponding to the three sub-figures in \fref{TangentialContact_example}, one can see that the normal load is in very good approximation constant over the $10$ tangential load cycles.
In particular, the normal load is invariant over the tangential load cycle; \ie, the variation is apparently below the standard deviation.
\\
The relative tangential displacement $\gT$ is shown in the bottom row of \fref{pN-time}.
The drift was mitigated as described in \ssref{instrumentation}.
Apparently, the motion is very smooth.
This verifies that the motion is not affected by undesired play, backlash or parasitic hysteresis, nor by unwanted natural dynamics of the rig.

\subsection{Verification of initial alignment of the nominally planar contact areas\label{sec:alignment}}
It is useful to recall here that the purpose of the \BMT is to test samples also in \emph{pristine condition}.
In other words, the surfaces should not have to undergo substantial wear before reliable normal load-indentation and frictional hysteresis cycles are acquired.
It is well-known that wear has a severe impact on those results.
We believe that the force-displacement relations under pristine condition are relevant for contacts that always remain in partial slip / partial liftoff, such as bolted or riveted joints, which are not expected to undergo substantial wear under normal operating conditions.
\\
The initial \emph{alignment} of the nominally flat pair of contact surfaces is a crucial aspect.
If there is an angular difference, \ie, the surface normal vectors are not collinear, contact would occur very localized (at a point or a line in the ideal case of smooth surfaces).
An unknown angular orientation hampers the interpretation of the acquired force-displacement relations, and, in particular, adds strong uncertainty to the tasks of identification/validation of contact models.
It is useful to recall the means to achieve good alignment from \ssref{SamplesAndHolder}:
tight tolerances, relatively small apparent contact area, finite rotation stiffness of the compliant mechanism.
\\
To verify the initial alignment, the contact was very briefly driven into gross slip.
The sole purpose of this was to generate visible traces of contact, in order to see whether the contact occurs in the complete apparent area, or only localized in a sub-region.
The gross slip phase was kept very brief in order to minimize the effect of wear on the alignment.
This is in full accordance with the aim of characterizing pristine surfaces.
It should also be emphasized that the \emph{\BMT in its present form is not qualified for well-defined gross slip operation}:
In the gross slip regime, the normal load rapidly decreases over a small number of cycles.
Also, the amplitude controller became unstable; due to the discontinuous stick-slip events, the  gross slip regime is more demanding from the control perspective, and no effort was made yet to obtain stable control there.
A gross slip upgrade of the \BMT is ongoing work.
\begin{figure}[h!]%
  \centering
  \def\svgwidth{0.66\textwidth}
\begingroup%
  \makeatletter%
  \providecommand\color[2][]{%
    \errmessage{(Inkscape) Color is used for the text in Inkscape, but the package 'color.sty' is not loaded}%
    \renewcommand\color[2][]{}%
  }%
  \providecommand\transparent[1]{%
    \errmessage{(Inkscape) Transparency is used (non-zero) for the text in Inkscape, but the package 'transparent.sty' is not loaded}%
    \renewcommand\transparent[1]{}%
  }%
  \providecommand\rotatebox[2]{#2}%
  \newcommand*\fsize{\dimexpr\f@size pt\relax}%
  \newcommand*\lineheight[1]{\fontsize{\fsize}{#1\fsize}\selectfont}%
  \ifx\svgwidth\undefined%
    \setlength{\unitlength}{510.23622047bp}%
    \ifx\svgscale\undefined%
      \relax%
    \else%
      \setlength{\unitlength}{\unitlength * \real{\svgscale}}%
    \fi%
  \else%
    \setlength{\unitlength}{\svgwidth}%
  \fi%
  \global\let\svgwidth\undefined%
  \global\let\svgscale\undefined%
  \makeatother%
  \begin{picture}(1,0.28301112)%
    \lineheight{1}%
    \setlength\tabcolsep{0pt}%
    \put(0,0){\includegraphics[width=\unitlength,page=1]{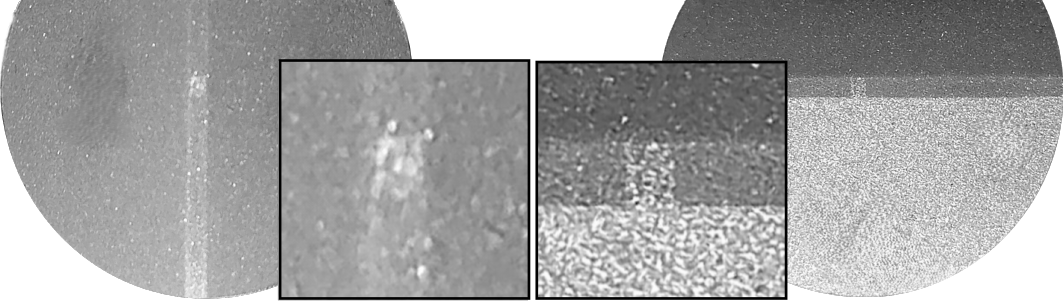}}%
    \put(0.35360285,0.03511356){\color[rgb]{0,0,0}\makebox(0,0)[lt]{\lineheight{1.25}\smash{\begin{tabular}[t]{l}$w_\mathrm{A}$\end{tabular}}}}%
    \put(0.42560992,0.12394287){\color[rgb]{0,0,0}\makebox(0,0)[lt]{\lineheight{1.25}\smash{\begin{tabular}[t]{l}$w_\mathrm{B}$\end{tabular}}}}%
    \put(0.59111983,0.0312497){\color[rgb]{0,0,0}\makebox(0,0)[lt]{\lineheight{1.25}\smash{\begin{tabular}[t]{l}$w_\mathrm{A}$\end{tabular}}}}%
    \put(0.67638963,0.11752743){\color[rgb]{0,0,0}\makebox(0,0)[lt]{\lineheight{1.25}\smash{\begin{tabular}[t]{l}$w_\mathrm{B}$\end{tabular}}}}%
    \put(0,0){\includegraphics[width=\unitlength,page=2]{FP_1_16_D_wear_V2.pdf}}%
    \put(0.12282327,0.25590275){\color[rgb]{0,0,0}\makebox(0,0)[lt]{\lineheight{1.25}\smash{\begin{tabular}[t]{l}upper sample\end{tabular}}}}%
    \put(0.74577221,0.25666447){\color[rgb]{0,0,0}\makebox(0,0)[lt]{\lineheight{1.25}\smash{\begin{tabular}[t]{l}lower sample\end{tabular}}}}%
  \end{picture}%
\endgroup%
  \caption{Photos of sample pair taken after driving the contact briefly into the gross slip regime. $w_{\mrm A}$, $w_{\mrm B}$ are the web thicknesses obtained from surface scanning prior to the test, \cf \fref{surfaceScan}.}
  \label{fig:homogeneous}
\end{figure}
\\
For the above reasons, the hysteresis cycles acquired in the brief gross slip regime were highly distorted and it does not appear valuable to show them here.
Still, to verify that the effect of wear is negligible, the energy $\int \FT \dd\gT$ was numerically evaluated over the complete time span from the activation of the shaker to its shutdown.
Hence, the contribution of both the relatively long partial slip operation as well as the very brief gross slip operation is accounted for.
This led to an estimated total energy over all cycles of $5~\mrm{mJ}$.
This value is deemed negligible.
For reference:
The energy required to wear down the initial misalignment and achieve complete contact of a similar apparent area with the Imperial College rig \cite{Schwingshackl.2012} amounts to a few hundred Joules (five orders of magnitude more).
\\
\fref{homogeneous} shows photos of the apparent contact areas after the described test.
The photos were obtained under regular light, whose angle was varied until the contact traces were sufficiently well visible.
The traces are distributed over the entire apparent contact area in a relatively uniform way.
From this, we conclude that the initial alignment of the surfaces achieved with the \BMT is sufficient, at least for the given surface roughness $\mrm{Ra}=1.6~\mum$, and the apparent area of about $1~\mrm{mm}\,\times\,1~\mrm{mm}$.

\section{Conclusions and outlook\label{sec:conclusions}}
The \BMT developed in the present work is useful for measuring frictional hysteresis cycles as well as normal load-indentation curves.
In particular, the \BMT is qualified to test nominally flat-on-flat contacts in the partial sliding regime.
The compliant mechanism, which represents the core of the \BMT, is regarded as key ingredient to achieve smooth and well-defined motion, which is a prerequisite for precise measurements in the sub-micrometer range.
The fact that the \BMT enables normal and tangential contact characterization, of the same pair of samples, without reassembly, from the pristine state of the surface, is expected to facilitate the validation of more predictive modeling approaches, which account for the relevant features of the surface topography (including roughness).
\\
Overall, the results obtained so far look very plausible.
Apart from the uncertainty in the normal displacement measurement, which has been attributed to viscous displacements within the sample holders, no major source of uncertainty has been identified.
More specifically, the remaining uncertainties are regarded as negligible compared to the variability inherent to frictional-unilateral contact behavior, expected for a sufficient number of nominally identical samples.
A more direct measurement of the indentation, which is not affected by the aforementioned viscous displacements, is planned.
Although the normal load was maintained well over the tangential load cycle, a feedback-controlled actuation of the normal load is regarded as useful upgrade.
A mid-term goal is to use the \BMT for setting up an open database of contact parameters for relevant material combinations, surface treatments and ambient conditions.

\section*{Acknowledgements}
This work has been funded jointly by the Federal Ministry for Economic Affairs and Climate Action (grant number 03EE5041G), MTU Aero Engines AG and Siemens Energy Global GmbH \& Co. KG.
The authors greatly acknowledge the financial support and the permission to publish.
The responsibility for the content of the publication rests authors.


\end{document}